\documentclass{PoS}

\usepackage{amsmath}
\usepackage{units}

\newcommand{\dd}{\ensuremath{\mathrm{d}}}

\newcommand{\Pgg}{\ensuremath{\mathrm{\gamma}}}
\newcommand{\Pp}{\ensuremath{\mathrm{p}}}
\newcommand{\Pgpz}{\ensuremath{\mathrm{\pi^0}}}
\newcommand{\Pgh}{\ensuremath{\mathrm{\eta}}}

\title{Measurement of double polarization observables in meson photoproduction off the proton with the CBELSA/TAPS experiment}

\ShortTitle{Measurement of double polarization observables with the CBELSA/TAPS experiment}

\author{\speaker{Jan Hartmann} \\%
        For the CBELSA/TAPS Collaboration\\
        HISKP, Bonn University\\
        E-mail: \email{hartmann@hiskp.uni-bonn.de}}


\abstract{
One of the remaining problems within the standard model is to gain a good understanding of the low energy regime of QCD, where perturbative methods fail. One key towards a better understanding is baryon spectroscopy. Unfortunately, in the past most baryon spectroscopy data have been obtained only using $\pi N$ scattering. To gain access to resonances with small $\pi N$ partial width, photoproduction experiments, investigating various final states, provide essential information. In order to extract the contributing resonances, partial wave analyses need to be performed. Here, the complete experiment is required to unambiguously determine the contributing amplitudes. This involves the measurement of carefully chosen single and double polarization observables. 

The Crystal Barrel/TAPS experiment with a longitudinally or transversely polarized target and an energy tagged, linearly or circularly polarized photon beam allows the measurement of a large set of polarization observables. Due to its good energy resolution, high detection efficiency for photons, and the nearly complete solid angle coverage, it is ideally suited for the measurement of the photoproduction of neutral mesons decaying into photons. 

Preliminary results for the target asymmetry $T$, recoil polarization $P$ and the double polarization observable $H$ are discussed for $\pi^{0}$ and $\eta$ photoproduction off the proton.}

\FullConference{XV International Conference on Hadron Spectroscopy \\
                 4-8/11/2013\\
                 Nara, Japan}

\begin{document}

\section{Introduction}
In order to extract the contributing resonances in photoproduction experiments, partial wave analyses need to be performed. A complete experiment is required to determine the contributing amplitudes. This involves the measurement of single and double polarization observables.

For single pseudoscalar meson photoproduction using a linearly polarized photon beam and a transversely polarized target, the cross section can be written in the form
\begin{equation}
\frac{\dd\sigma}{\dd\Omega} = \left(\frac{\dd\sigma}{\dd\Omega}\right)_0 \left[1-P_{\Pgg}\Sigma\cos(2\phi) - P_xP_{\Pgg}H\sin(2\phi) - P_y\left(P_{\Pgg}P\cos(2\phi)-T\right)\right]
\end{equation}
where $\left(\frac{\dd\sigma}{\dd\Omega}\right)_0$ is the unpolarized cross section, $\Sigma$, $T$, $P$, and $H$ are the occurring polarization observables \cite{barker:1975}, $P_{\Pgg}$ is the degree of linear photon polarization, and $\phi$ is the azimuthal angle of the photon polarization plane with respect to the reaction plane. $P_x$ and $P_y$ are the degrees of target polarization in the reaction plane and perpendicular to it, respectively.

In this paper, preliminary new data on the observables $T$, $P$, and $H$ will be presented for the reactions $\vec{\Pgg}\,\vec{\Pp} \to \Pp \Pgpz$ and $\vec{\Pgg}\,\vec{\Pp} \to \Pp \Pgh$, thus providing an important step towards the complete experiment.

\section{Data analysis}
The data presented have been obtained with the CBELSA/TAPS experiment at ELSA \cite{hillert:2006}. The detector system consists of two electromagnetic calorimeters, the Crystal Barrel \cite{aker:1992} and the MiniTAPS detector \cite{novotny:1991}, together covering the polar angle range from $1^\circ$ to $156^\circ$ and the full azimuthal angle. For charged particle identification, a three-layer scintillating fiber detector \cite{suft:2005} surrounding the target, and plastic scintillators in forward direction were used. The frozen spin butanol target \cite{bradtke:1999} was operated with a superconducting saddle coil providing a homogeneous magnetic field perpendicular to the beam direction, reaching an average target polarization of $\unit[74]{\%}$. Data have been taken with two opposite settings of the target polarization direction (named $\uparrow$ and $\downarrow$).

For this analysis, a data set obtained with a primary electron energy of $\unit[3.2]{GeV}$ was used. The energy tagged photon beam was linearly polarized by means of coherent bremsstrahlung \cite{elsner:2009} with a maximum polarization of $\unit[65]{\%}$ at $E_\gamma = \unit[850]{MeV}$. Two perpendicular settings of the polarization plane were used (named $\parallel$ and $\perp$). 

The data sample was selected for events with three distinct calorimeter hits. First all three are treated as photon candidates, three $2\gamma$ invariant masses are formed and a cut on the $\gamma\gamma$ invariant mass is applied to select the reactions $\vec{\Pgg}\,\vec{\Pp} \to \Pp \Pgpz \to \Pp\Pgg\Pgg$ and $\vec{\Pgg}\,\vec{\Pp} \to \Pp \Pgh \to \Pp\Pgg\Pgg$. Then, with the remaining calorimeter hit as the proton candidate, additional cuts to ensure energy and momentum conservation are applied. This results in a final event sample containing a total of $1.4$ million $\Pp\Pgpz$ and $140000$ $\Pp\Pgh$ events with a background contribution of $<\unit[1]{\%}$ and $<\unit[2]{\%}$, respectively. The selected events for each of the four combinations of beam and target polarization directions were normalized w.r.t. the average beam and target polarization and the integrated photon flux.

The butanol target contains unpolarized C and O nuclei. The dilution factor $d$ takes into account that photoproduction of $\Pgpz$ or $\Pgh$ off unpolarized, bound nucleons in C or O nuclei cannot be completely separated from the reactions on the polarized, free protons. $d$ is a function of $E_\gamma$ and $\cos\theta$, which is determined using data for which the butanol target was replaced by a carbon foam target inside the cryostat.

\begin{figure}
\centering
\hfill
\includegraphics[width=.3\textwidth]{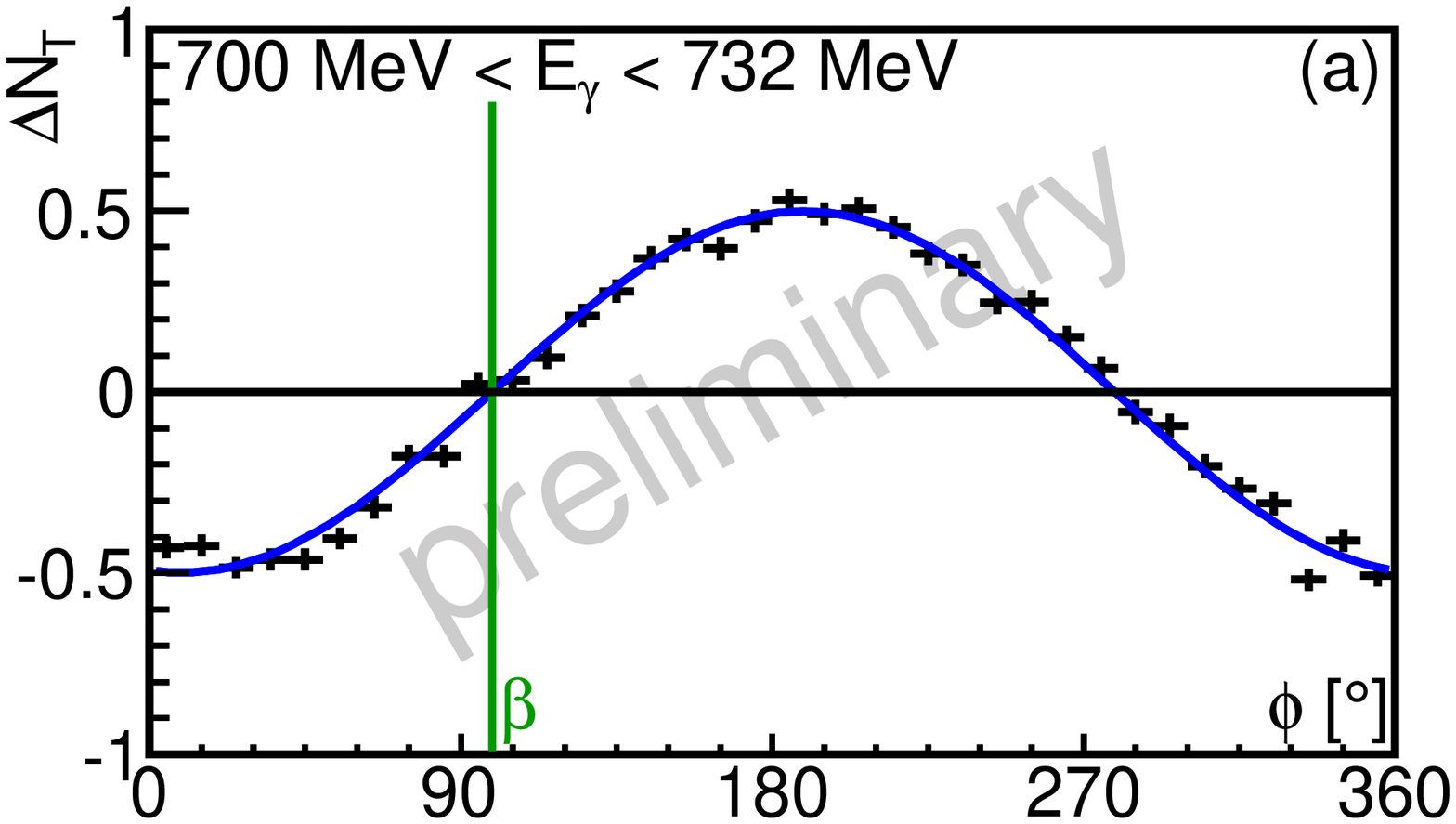} \hfill
\includegraphics[width=.3\textwidth]{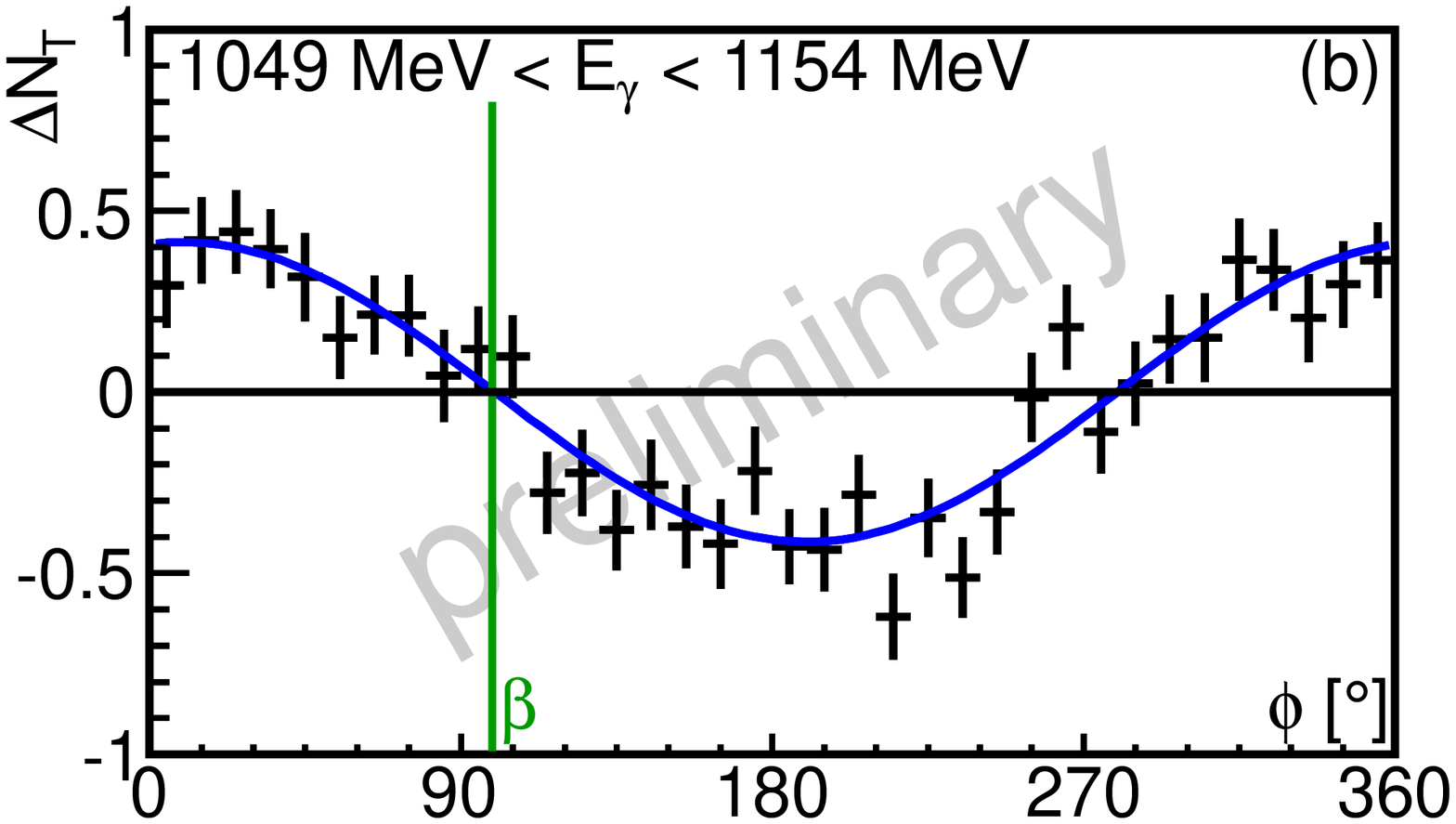} \hfill
\includegraphics[width=.3\textwidth]{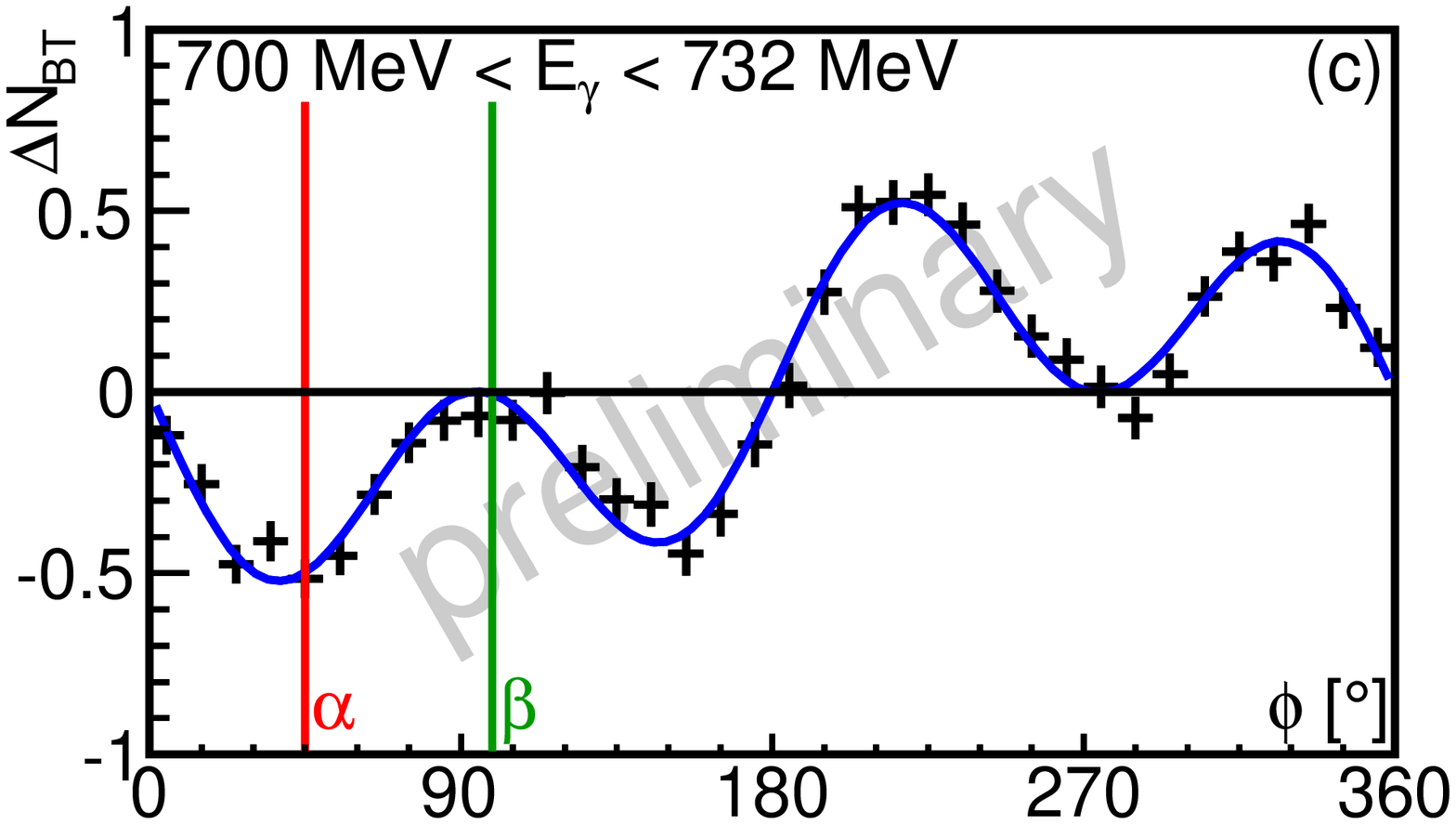} \hfill
\caption{Examples for measured azimuthal yield asymmetries (integrated over the polar angle) used to extract the target asymmetry $T$ in $\Pgpz$ (a) and $\Pgh$ photoproduction (b), and $P$ and $H$ in $\Pgpz$ photoproduction (c).}
\label{fig:phidistr}
\end{figure}

Since the detector acceptance is identical for all polarization settings, the cross section can be replaced by the normalized yield $N$, and the target asymmetry $T$ is determined from a fit to the azimuthal yield asymmetry
\begin{equation}
\Delta N(\phi)_\text{\,T} = \frac{1}{d \cdot P_t} \cdot \frac{N_\uparrow-N_\downarrow}{N_\uparrow+N_\downarrow} = T \cdot \sin(\beta-\phi) ;\qquad d(E_{\Pgg},\theta) = \frac{N_\text{butanol} - N_\text{carbon}}{N_\text{butanol}}
\end{equation}
with average target polarization $P_t$ and $\beta=100^\circ$ being the direction of the target polarization in the $\uparrow$ setting. 
A typical example for such a fit is shown in Fig.~\ref{fig:phidistr}\,(a) and Fig.~\ref{fig:phidistr}\,(b). Two additional observables are accessible by also considering the linear polarization of the photon beam. In addition to the observable $H$, one also has access to the recoil polarization $P$, without the difficulties associated with a direct measurement of the recoil proton polarization. In order to extract both observables from the data, all four combinations of beam and target polarization settings are used:
\begin{align}
\Delta N(\phi)_\text{\,BT} &=
\frac{1}{d \cdot P_{\Pgg}\,P_t} \cdot \frac{(N_{\perp\uparrow}-N_{\perp\downarrow})-(N_{\parallel\uparrow}-N_{\parallel\downarrow})}{(N_{\perp\uparrow}+N_{\perp\downarrow})+(N_{\parallel\uparrow}+N_{\parallel\downarrow})} \\
&= P \sin(\beta-\phi)\cos(2(\alpha-\phi)) + H \cos(\beta-\phi)\sin(2(\alpha-\phi)) \nonumber
\end{align}
with average beam polarization $P_{\Pgg}$ and $\alpha=45^\circ$ being the direction of the polarization plane in the $\parallel$ setting. Again, the observables are determined by a fit to the azimuthal yield asymmetry, as shown by the example in Fig.~\ref{fig:phidistr}\,(c).

\section{Preliminary results}
\begin{figure}
\includegraphics[width=\textwidth]{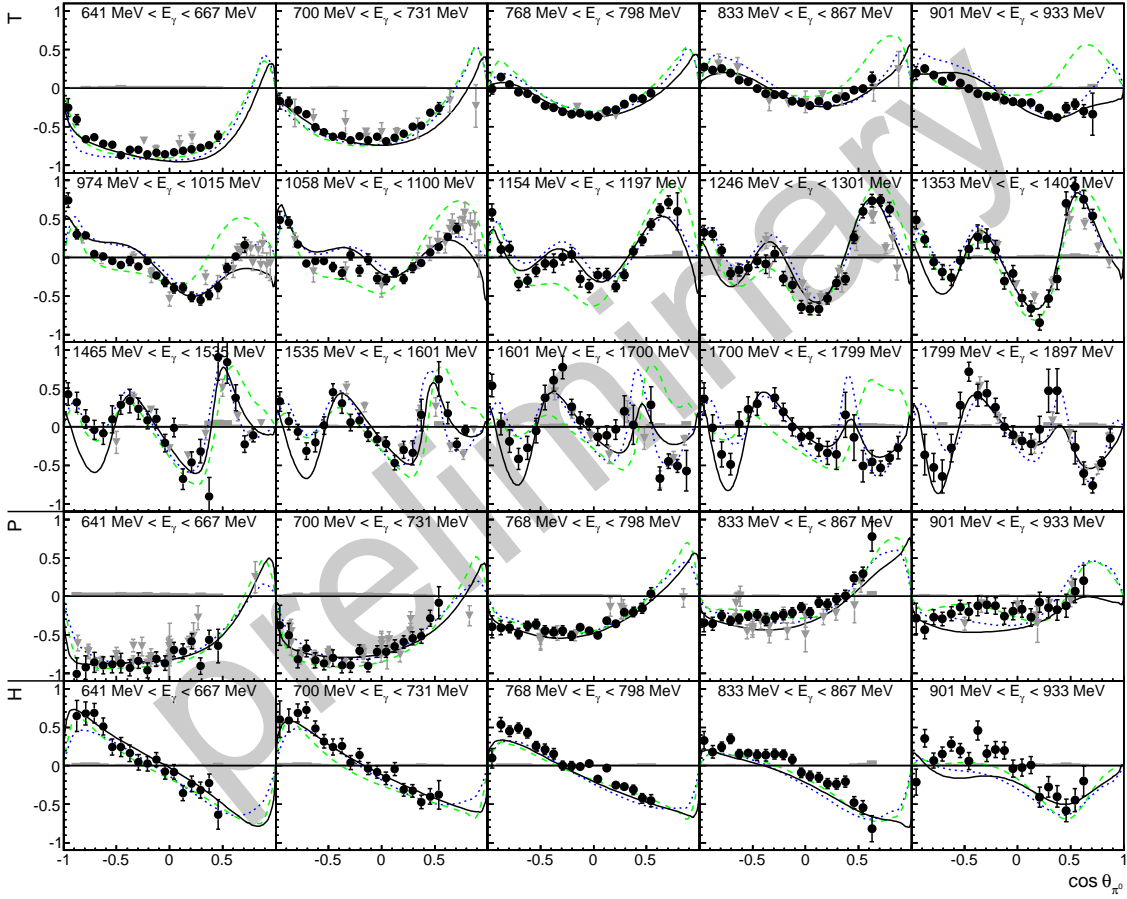}
\caption{Preliminary distributions of $T$ (top 3 rows), $P$ (4th row), and $H$ (bottom row) for the reaction $\vec{\Pgg}\,\vec{\Pp}\to\Pp\Pgpz$ (black) as a function of the $\Pgpz$ CMS angle, compared to previous measurements \cite{Olddata} of $T$ and $P$ (gray) and the predictions of the BnGa \cite{bnga} (solid), SAID \cite{said:2012} (dotted), and MAID \cite{maid:2007} (dashed) analyses. Only every second energy bin is shown here.}
\label{fig:pi0}
\end{figure}
\begin{figure}
\includegraphics[width=\textwidth]{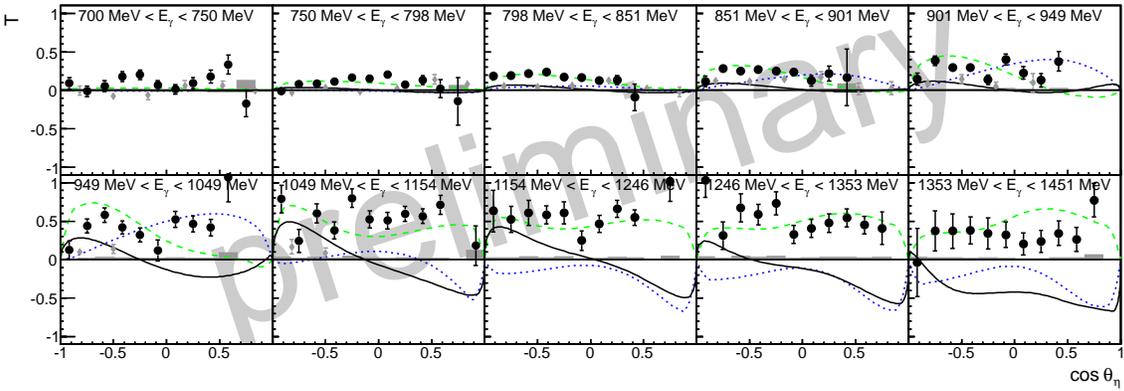}
\caption{Very preliminary distributions of $T$ for the reaction $\Pgg\,\vec{\Pp}\to\Pp\Pgh$ (black) as a function of the $\Pgh$ CMS angle, compared to previous measurements \cite{bock:1998} (gray) and the predictions of the BnGa \cite{bnga} (solid), SAID \cite{said:2009} (dotted), and MAID \cite{maid:2007} (dashed) analyses.}
\label{fig:eta}
\end{figure}

Preliminary results for the observables $T$, $P$, and $H$ are shown in Fig.~\ref{fig:pi0} for $\Pgpz$ photoproduction. The agreement with previous measurements of $T$ performed at the Daresbury $\unit[5]{GeV}$ electron synchrotron \cite{booth:1977} is quite good. If compared to the predictions of different state of the art PWA, discrepancies are observed between the different predictions and the data, especially at higher energies. This indicates a not yet satisfactory understanding of $\Pgpz$ photoproduction.

Very preliminary results for the target asymmetry $T$ in $\Pgh$ photoproduction are shown in Fig.~\ref{fig:eta}. The results seem to be inconsistent with previous measurements done at ELSA \cite{bock:1998}, but a detailed study of the systematic uncertainties needs to be done before a final conclusion can be drawn. A comparison of the PWA predictions to the data shows significant discrepancies. This data will without doubt contribute to a better understanding of $\Pgh$ photoproduction. It provides an important next step towards the complete experiment.

\section{Summary}
Data have been taken with the CBELSA/TAPS experiment using the newly developed transversely polarized target and a linearly polarized photon beam. The preliminary results show the excellent quality of the data. Further measurements are planned to increase the statistics to investigate other reactions and higher energies.
Together with the measurements with a longitudinally polarized target and both linearly \cite{thiel:2012} or circularly \cite{gottschall:2014} polarized photon beams this is an important step towards the complete experiment and will provide further constraints for the partial wave analysis.

This work was supported by the \textit{Deutsche Forschungsgemeinschaft} within SFB/TR-16 and the \textit{Schweizerischer Nationalfonds}.

\bibliographystyle{JHEP}

\bibliography{cb_hartmann}

\providecommand{\href}[2]{#2}\begingroup\raggedright\begin{thebibliography}{10}

\bibitem{barker:1975}
I.~S. Barker, A.~Donnachie, and J.~K. Storrow, {\it Complete experiments in
  pseudoscalar photoproduction},  {\em Nucl. Phys.} {\bf B95} (1975) 347.

\bibitem{hillert:2006}
W.~Hillert, {\it {The Bonn Electron Stretcher Accelerator ELSA}},  {\em Eur.
  Phys. J. A} {\bf 28} (2006) 139.

\bibitem{aker:1992}
E.~Aker et~al., {\it {The Crystal Barrel spectrometer at LEAR}},  {\em Nucl.
  Instr. Methods A} {\bf 321} (1992) 69.

\bibitem{novotny:1991}
R.~Novotny, {\it {The BaF$_2$ photon spectrometer TAPS}},  {\em IEEE Trans.
  Nucl. Sci.} {\bf NS-38} (1991) 379.

\bibitem{suft:2005}
G.~Suft et~al., {\it {A scintillating fibre detector for the Crystal Barrel
  experiment at ELSA}},  {\em Nucl. Instr. Methods A} {\bf 531} (2005) 416.

\bibitem{bradtke:1999}
C.~Bradtke, H.~Dutz, H.~Peschel, et~al., {\it {A new frozen-spin target for
  4$\pi$ particle detection}},  {\em Nucl. Instr. Methods A} {\bf 436} (1999)
  430.

\bibitem{elsner:2009}
D.~Elsner et~al., {\it {Linearly polarised photon beams at ELSA}},  {\em Eur.
  Phys. J. A} {\bf 39} (2009) 373.

\bibitem{Olddata}
W.~Briscoe, D.~Schott, I.~Strakovsky, and R.~Workman, {\it {GWU analysis
  center}},  {\href{http://gwdac.phys.gwu.edu}{\tt http://gwdac.phys.gwu.edu}}.

\bibitem{bnga}
{\it {Bonn-Gatchina partial wave analysis, solution BG2011-02}},
  {\href{http://pwa.hiskp.uni-bonn.de}{\tt http://pwa.hiskp.uni-bonn.de}}.

\bibitem{said:2012}
R.~Workman, M.~Paris, W.~Briscoe, and I.~Strakovsky, {\it {Unified
  Chew-Mandelstam SAID analysis of pion photoproduction data}},  {\em Phys.
  Rev. C} {\bf 86} (2012) 015202.

\bibitem{maid:2007}
D.~Drechsel, S.~S. Kamalov, and L.~Tiator, {\it {Unitary Isobar Model -
  MAID2007}},  {\em Eur. Phys. J. A} {\bf 34} (2007) 69.

\bibitem{bock:1998}
A.~Bock et~al., {\it {Measurement of the Target Asymmetry of $\eta$ and $\pi^0$
  Photoproduction on the Proton}},  {\em Phys. Rev. Lett.} {\bf 81} (1998) 534.

\bibitem{said:2009}
M.~Dugger et~al., {\it {$\pi{}^{+}$ photoproduction on the proton for photon
  energies from 0.725 to 2.875 GeV}},  {\em Phys. Rev. C} {\bf 79} (2009)
  065206.

\bibitem{booth:1977}
P.~Booth et~al., {\it {The polarised target asymmetry for neutral pion
  photoproduction from protons}},  {\em Nucl. Phys.} {\bf B121} (1977) 45.

\bibitem{thiel:2012}
A.~Thiel et~al., {\it {Well-established nucleon resonances revisited by
  double-polarization measurements}},  {\em Phys. Rev. Lett.} {\bf 109} (2012)
  102001.

\bibitem{gottschall:2014}
M.~Gottschall et~al., {\it {First Measurement of the Helicity Asymmetry for
  $\gamma p \to p \pi^0$ in the Resonance Region}},  {\em Phys. Rev. Lett.}
  {\bf 112} (2014) 012003.

\end{thebibliography}\endgroup

\end{document}